\documentclass[10pt, conference]{IEEEtran} 

\usepackage{url}
\usepackage{amsfonts}
\usepackage{amssymb}
\usepackage{amsmath}
\usepackage{graphicx}
\usepackage{epsfig}
\usepackage{bm} 
\usepackage{bbm} 
\usepackage{enumerate}

\newcommand{\beq}[1]{\begin{equation}\label{#1}}
\newcommand{\eeq}{\end{equation}}

\newcommand{\beqn}[1]{\begin{eqnarray}\label{#1}}
\newcommand{\eeqn}{\end{eqnarray}}

\newtheorem{thmbody}{Theorem}
\newenvironment{thm}{
\begin{thmbody}
	}{
	\end{thmbody} 
	}
\newtheorem{dfnbody}{Definition}

\newtheorem{corbody}{Corollary}

\newtheorem{lemmabody}{Lemma}
\newenvironment{lemma}{
\begin{lemmabody}
	}{
	\end{lemmabody} 
	}
\newtheorem{propbody}{Proposition}

\newenvironment{proof}{
	{\it Proof:}
	}{
 $\Box$
	}

\usepackage{dsfont}
\begin{document}
\title{Stochastic Interpretation for the Arimoto Algorithm}


\author{\IEEEauthorblockN{Sergey Tridenski}
\IEEEauthorblockA{EE - Systems Department\\Tel Aviv University\\
Tel Aviv, Israel\\
Email: sergeytr@post.tau.ac.il}
\and
\IEEEauthorblockN{Ram Zamir}
\IEEEauthorblockA{EE - Systems Department\\Tel Aviv University\\
Tel Aviv, Israel\\
Email: zamir@eng.tau.ac.il}}



\maketitle
\begin{abstract}
The Arimoto algorithm computes the Gallager function
$\max_Q {E\mathstrut}_{0}^{}(\rho,Q)$
for a given channel ${P\mathstrut}_{}^{}(y \,|\, x)$ and parameter $\rho$, by means
of alternating maximization. Along the way, it generates a sequence
of input distributions ${Q\mathstrut}_{1}^{}(x)$, ${Q\mathstrut}_{2}^{}(x)$, ... ,
that converges to the maximizing
input ${Q\mathstrut}_{}^{*}(x)$.  
We propose a stochastic interpretation for the Arimoto algorithm.
We show that for a random (i.i.d.) codebook with
a distribution ${Q\mathstrut}_{k}^{}(x)$, the next distribution ${Q\mathstrut}_{k+1}^{}(x)$ in the Arimoto
algorithm is equal to the type (${Q\mathstrut}'$) of the feasible transmitted codeword
that maximizes the conditional Gallager exponent (conditioned on a specific
transmitted codeword type ${Q\mathstrut}'$). This interpretation is a first step
toward finding a stochastic mechanism for on-line channel input adaptation.
\footnote{This work of S. Tridenski and R. Zamir was supported in part by the Israeli Science Foundation (ISF), grant \# 870/11}
\end{abstract}



%
%
%
%


\section{Introduction} \label{Introduction}
Optimization of the transmitter output, under a set of constraints and channel conditions, is an important goal in digital communication. In information theoretic terms, it corresponds to selecting the channel input distribution ${Q\mathstrut}_{}^{*}$
that maximizes the mutual information (to achieve the Shannon capacity $C$ of the channel) or the Gallager
function (to achieve the best error exponent $E(R)$).
Since the channel conditions are often changing, practical schemes tend to gradually adapt to these conditions, using feedback from the receiver to the transmitter.

A previous work \cite{ZamirRose01}, \cite{KochmanZamir02} found the phenomenon of {\em natural type selection}
(NTS) in adaptive lossy source coding: the empirical distribution (or {\em type})
${Q\mathstrut}'$ of the first ``$D$-matching'' codeword in an ordered random codebook,
generated i.i.d. according to a distribution $Q$,
amounts asymptotically (as the word length goes to infinity) to a single iteration (from $Q$ to ${Q\mathstrut}'$) of the Blahut algorithm for rate-distortion function computation \cite{Blahut72}. By iterating the NTS step, the sequence of codebook
distributions $Q$, ${Q\mathstrut}'$, ${Q\mathstrut}'', \ldots$ converges to the reproduction distribution ${Q\mathstrut}_{}^{*}$ that realizes
the rate-distortion function $R(D)$. This result gives a stochastic flavor to the Blahut algorithm, by interpreting each iteration as an instance of the conditional limit theorem of large deviation theory.

In this paper we provide a first step toward the channel-coding counterpart of NTS.
The underlying idea is that the type of a ``good'' codeword (good in the sense of having a low individual error probability) indicates the direction to which the codebook distribution should evolve.
Similarly to the source coding case,
the type of the ``best'' codeword (in the above sense) amounts to a single iteration
of the Arimoto error-exponent computation algorithm \cite{Arimoto76}.
Thus, by iteration of the NTS step we obtain
a sequence of input distributions, $Q$, ${Q\mathstrut}'$, ${Q\mathstrut}'', \ldots$,
that converges to $Q^*$,
the input that maximizes the Gallager function ${E\mathstrut}_{0}^{}(\rho,Q)$.
By a proper selection of the parameter $\rho$,
this input leads to the optimum random-coding error exponent
$E(R)$ at a coding rate $R$,
and to the channel capacity $C$ in the limit as $\rho \rightarrow 0$.

Section~\ref{Proposition} gives some background on the Arimoto algorithm
and presents our main result. Section~\ref{Proof} proves the main result. Discussion is given in Section~\ref{Discussion}.


\section{Background and Statement of Result} \label{Proposition}
In our search for a dual phenomenon in channel coding
we consider an alternating maximization algorithm
for random coding exponent computation,
discovered by Arimoto in 1976 \cite{Arimoto76}.
This algorithm is a refinement of Arimoto's earlier work from 1972 on
capacity computation \cite{Arimoto72} and Blahut's work
\cite{Blahut72} on rate-distortion function (and capacity) computation.


\subsection{\bf Capacity and Error Exponent}
\label{CapacityErrorExponent}
Consider a discrete memoryless channel, with a transition probability
distribution $P(y \,|\, x)$ from the input $x$ to the output $y$.
Let $Q(x)$ denote an arbitrary probability assignment on the channel input alphabet.
It is well known, \cite{Gallager68,CoverBook},
that the capacity $C$ of this channel is given
by maximizing the mutual information $I(Q,P)$ over the input $Q$:
\begin{equation} \label{eqCapacity}
C = \max_Q I(Q,P) ,
\end{equation}
where
\begin{equation} \label{eqMutualInformation}
I(Q,P) =
\sum_{y}^{}\sum_{x}^{}Q(x){P\mathstrut}_{}^{}(y \,|\, x)
\log \bigg[\frac{P(y \,|\, x)} {\sum_{x'} Q(x')P(y \,|\, x') } \bigg] .
\end{equation}
Furthermore, for any coding rate $R < I(Q,P)$,
if we generate $M = 2^{nR}$ length-$n$ codewords
by an i.i.d. distribution $Q$,
then the error probability in maximum likelihood decoding is bounded from above by
\begin{equation} \label{eqPeBound}
P_{\rm e} \leq \exp \big\{-n\big[E_{0}^{}\big(\rho, Q\big) -\rho {R\mathstrut}_{}^{} \big] \big\},
\end{equation}
for all $0 \leq \rho \leq 1$, where
%
%
\begin{multline} \label{eqGallagerFunction}
E_{0}^{}(\rho,\,Q) \; = \;
-\log \sum_{y}^{}\bigg[\sum_{x}^{}Q(x){P\mathstrut}_{}^{\frac{1}{1+\rho}}(y \,|\, x)\bigg]^{1+\rho}
\end{multline}
is the Gallager function defined in \cite[eq.~(5.6.14)]{Gallager68}.
%

The optimum value of the parameter $\rho$ that maximizes the error exponent
in (\ref{eqPeBound}) decreases as the rate $R$ increases,
and vanishes as $R$ approaches $I(Q,P)$.
In particular,
$I(Q,P) = \lim_{\rho \rightarrow 0} E_0(\rho,Q) / \rho$.
%
For a fixed value of the parameter $\rho$,
the channel input distribution 
that maximizes the Gallager function (\ref{eqGallagerFunction}),
\begin{equation} \label{eqQ*}
Q^*_\rho
=
\arg\max_Q E_0(\rho,Q) ,
\end{equation}
maximizes also the error exponent in (\ref{eqPeBound}).\footnote
{
For a non-redundant channel input alphabet,
the optimum input $Q^*_\rho$ is unique
\cite[chap.~4.5, corr.~3]{Gallager68}.
}
It follows that $Q^*_\rho \rightarrow Q^*_C$, as $\rho \rightarrow 0$,
where
$Q^*_C$ is the capacity-achieving input distribution (\ref{eqCapacity}).

Unfortunately, however,
except for symmetric channels (where the optimum input is uniform
\cite{Gallager68}), these optimum distributions do not have a closed-form solution.
This motivates the Arimoto and Blahut iterative algorithms
\cite{Arimoto72,Blahut72,Arimoto76}.


\subsection{\bf Arimoto's Alternating Maximization} 
\label{ArimotoAlgorithm}
Instead of maximizing $E_{0}^{}(\rho,\,Q)$ directly over $Q$ (which may be a hard task),
the Arimoto algorithm alternates between two maximization steps.
To this end, the Gallager function
(\ref{eqGallagerFunction}) can be rewritten as \cite{Arimoto76}
\begin{multline}
-\log \sum_{y}^{}\bigg[\sum_{x}^{}Q(x){P\mathstrut}_{}^{\frac{1}{1+\rho}}(y \,|\, x)\bigg]^{1+\rho}
\; = \; \\
= \; -\log \sum_{y}^{}\sum_{x}^{}Q(x){P\mathstrut}_{}^{}(y \,|\, x)\bigg[\frac{{\Phi\mathstrut}_{}^{}(x \,|\, y)}{Q(x)}\bigg]^{-\rho},
\label{eqArimotoForm}
\end{multline}
where ${\Phi\mathstrut}_{}^{}(x \,|\, y)$ achieving the equality in  (\ref{eqArimotoForm}) is a transition probability matrix given by
\begin{equation} \label{eqFrozen}
{\Phi\mathstrut}_{}^{}(x \,|\, y) \; \triangleq \;
\frac{Q(x){P\mathstrut}_{}^{\frac{1}{1+\rho}}(y \,|\, x)}
{\sum_{x'}^{}Q(x'){P\mathstrut}_{}^{\frac{1}{1+\rho}}(y \,|\, x')}.
\end{equation}
If we keep ${\Phi\mathstrut}_{}^{}(x \,|\, y)$ ``frozen''
and maximize the right hand side of (\ref{eqArimotoForm})
with respect to the probability vector $Q$,
the maximizing distribution is given by \cite{Arimoto76}
\begin{equation} \label{eqNewQ1}
{Q\mathstrut}'(x) \; = \; \frac{1}{K}\bigg[\sum_{y}^{}{P\mathstrut}_{}^{}(y \,|\, x){\Phi\mathstrut}_{}^{-\rho}(x \,|\, y)\bigg]_{}^{-1/\rho},
\end{equation}
where $K$ is the normalizing constant that keeps $\sum_{x}^{}{Q\mathstrut}'(x)=1$.
The expression (\ref{eqFrozen}) now contains the ``previous'' distribution $Q$.
Plugging (\ref{eqFrozen}) into (\ref{eqNewQ1}) we get
a recursive expression for the new distribution in terms of the previous one:
\begin{align}
& {Q\mathstrut}'(x) \; =
\nonumber \\
& \frac{1}{K}\,Q(x)\bigg[\sum_{y}^{}{P\mathstrut}_{}^{\frac{1}{1+\rho}}(y \,|\, x)\bigg(\sum_{x'}^{}Q(x'){P\mathstrut}_{}^{\frac{1}{1+\rho}}(y \,|\, x')\bigg)^{\rho}\,\bigg]_{}^{-1/\rho}.
\label{eqNewQ2}
\end{align}

At this point we have completed only one of the two maximization steps of the Arimoto algorithm and it is still unclear whether the new distribution ${Q\mathstrut}'$, determined by (\ref{eqNewQ2}), does not result in a lower value of the Gallager function $E_{0}^{}(\rho,\,{Q\mathstrut}')$.
As a next step,
we observe that the right hand side expression in the definition (\ref{eqFrozen}) has an extra meaning of the maximizer of (\ref{eqArimotoForm})
as a function of a dummy transition probability matrix ${\Phi\mathstrut}_{}^{}(x \,|\, y)$.
Therefore, replacing $Q$ with ${Q\mathstrut}'$
in both (\ref{eqFrozen}) and (\ref{eqArimotoForm})
we get $E_{0}^{}(\rho,\,{Q\mathstrut}')$ and ascertain that
\begin{displaymath} \label{eqOneIteration}
E_{0}^{}(\rho,\,{Q\mathstrut}') \; \geq \; E_{0}^{}(\rho,\,Q).
\end{displaymath}
Furthermore, as shown in \cite{Arimoto76}, an iterative application of the
recursion (\ref{eqNewQ2}),
starting from an arbitrary (nonzero) probability vector ${Q\mathstrut}_{1}^{}$,
produces a monotonically converging sequence
$\;E_{0}^{}(\rho,\,{Q\mathstrut}_{k}^{})\,\nearrow\,\max_{Q}^{}E_{0}^{}(\rho,\,{Q\mathstrut}_{}^{})$
and
$Q_k \rightarrow Q^*_\rho$,
as $k \rightarrow \infty$.

Note that the limit of (\ref{eqNewQ2}) as $\rho \rightarrow 0$
corresponds to computing the capacity $C$.
Specifically,
by L'Hopital's rule, (\ref{eqNewQ2}) becomes
\begin{align}
& {Q\mathstrut}'(x) \; =
\nonumber \\
& \frac{1}{K} \, Q(x)
\exp \Bigg\{
\sum_{y}^{}
{P\mathstrut}(y \,|\, x)
\log\,\frac{{P\mathstrut}(y \,|\, x)}{ \sum_{x'}^{}Q(x'){P\mathstrut}(y \,|\, x') }
\Bigg\}.
\label{eqNewQ2_capacity}
\end{align}
which is one iteration in computing the capacity-achieving distribution
in Arimoto's original algorithm from 1972 \cite{Arimoto72}.

Our goal is to find a stochastic mechanism which produces a new type of the form (\ref{eqNewQ2}).
Gallager's upper bound on error probability seems to be a relevant tool for this purpose.


\subsection{{\bf Conditional Gallager Exponent for a Given Codeword}} \label{GallagerBound}
Suppose that $M$ messages are used to communicate through an $n$-dimensional channel
${P\mathstrut}_{}^{}({\bf y}\,|\,{\bf x})$ (not necessarily memoryless).
Each message $m$, $1 \leq m \leq M$, is represented by a codeword ${\bf x}_{m}^{}$ of length $n$, selected independently
with the probability measure $Q({\bf x})$.

Let ${\bf x}_{m}^{}$ be a fixed channel input selected to represent a certain message $m$.
Given this ${\bf x}_{m}^{}$, consider the probability to get an error at the decoder, after we have independently generated (according to $Q({\bf x})$) the codewords for each of the $M-1$ alternative messages $m' \neq m $
and have sent
the message $m$ through the channel. This probability, denoted ${\overline{\!P\mathstrut}}_{{\rm e},\,{\bf x}_{m}^{}}$, can be bounded from above as
\begin{equation} \label{eqEnsemble}
{\overline{\!P\mathstrut}}_{{\rm e},\,{\bf x}_{m}^{}} \; \leq \; {M\mathstrut}_{}^{\rho}\sum_{\bf y}{P\mathstrut}_{}^{1-s\rho}({\bf y}\,|\,{\bf x}_{m}^{})\bigg[\sum_{{\bf x}}^{}Q({\bf x}){P\mathstrut}_{}^{s}({\bf y}\,|\,{\bf x})\bigg]_{}^{\rho},
\end{equation}
where $s > 0$ and $0 \leq \rho \leq 1$.
This is the Gallager bound
\cite[eq.~(5.6.10)]{Gallager68},
except for the averaging over the transmitted word ${\bf x}_{m}^{}$.

For the rest of the paper, we assume that both the probability measure $Q({\bf x})$
and the channel ${P\mathstrut}_{}^{}({\bf y}\,|\,{\bf x})$ are memoryless:
\begin{displaymath}
Q({\bf x})= \prod_{k=1}^{n}Q(x_{k}^{}), \;\;\;\;\;\; {P\mathstrut}_{}^{}({\bf y}\,|\,{\bf x})\; = \; \prod_{k=1}^{n}P(y_{k}^{} \,|\, x_{k}^{}).
\end{displaymath}
In this case, the conditional bound (\ref{eqEnsemble})
depends only on the type of the fixed word ${\bf x}_{m}^{}$,
and
takes the form shown by the following lemma, which is proved in the Appendix:


\vspace{3mm}

\begin{lemma} [Conditional Gallager exponent] \label{lemma0}
{\em For memoryless $Q({\bf x})$ and ${P\mathstrut}_{}^{}({\bf y}\,|\,{\bf x})$,
and a transmitted codeword ${\bf x}_{m}^{}$,
\begin{equation} \label{eqExpMemoryless}
{\overline{\!P\mathstrut}}_{{\rm e},\,{\bf x}_{m}^{}} \; \leq \;
\exp \big\{-n\big[E_{0}^{}\big(s, \rho, Q, {Q\mathstrut}_{{\bf x}_{m}^{}}^{}\big) -\rho {R\mathstrut}_{}^{} \big] \big\},
\end{equation}
where ${R\mathstrut}_{}^{} \, \triangleq \, \frac{1}{n}\log M$ is the coding rate,
$Q_{{\bf x}_{m}^{}}^{}$ is the type of ${\bf x}_{m}^{}$,
and}
\begin{multline} \label{eqExp}
E_{0}^{}(s, \rho, Q, \widetilde{Q}) \; \triangleq \;\\
-\sum_{x}^{} \widetilde{Q}(x)\log \sum_{y}^{}{P\mathstrut}_{}^{1-s\rho}(y \,|\, x)\bigg[\sum_{x'}^{}Q(x'){P\mathstrut}_{}^{s}(y \,|\, x')\bigg]^{\rho}.
\end{multline}
\end{lemma}

\vspace{3mm}


It can be shown that
the bound (\ref{eqExpMemoryless}) can be exponentially
tight 
for a certain choice of the parameters $s$ and $\rho$,
which is determined by the type $Q_{{\bf x}_{m}^{}}^{}$ and the code rate ${R\mathstrut}_{}^{}$.
%
Nevertheless, to enhance the relation to Gallager's analysis \cite{Gallager68},
we shall restrict attention to the (generally suboptimal) choice of $s = \frac{1}{1+\rho}$,
and use the simplified notation
\begin{multline}
\label{eqSimplifiedExponent}
E_{0}^{}(\rho, Q, \widetilde{Q})
\, \triangleq \,
E_{0}^{}\bigl(s=\tfrac{1}{1+\rho}, \,\rho, \,Q, \widetilde{Q}\big) =
\\
-\sum_{x}^{} \widetilde{Q}(x)\log \sum_{y}^{}{P\mathstrut}_{}^{\frac{1}{1+\rho}}(y \,|\, x)
\bigg[\sum_{x'}^{}Q(x'){P\mathstrut}_{}^{\frac{1}{1+\rho}}(y \,|\, x')\bigg]^{\rho}.
\end{multline}

The conditional Gallager function
$E_{0}^{}\big(\rho, Q, {Q\mathstrut}_{{\bf x}_{m}^{}}^{}\big)$
plays a similar role to the
Gallager function $E_{0}^{}(\rho,\,Q)$,
conditioned on a transmitted codeword of type $Q_{{\bf x}_{m}^{}}^{}$.
In fact, it is easy to show from (\ref{eqGallagerFunction})
and (\ref{eqSimplifiedExponent}) that
\begin{equation} \label{eqRelation2Gallager}
E_{0}^{}(\rho,\,Q)
=
\min_{\tilde{Q}} \Big\{ E_{0}^{}(\rho, Q, \widetilde{Q}) + D(\widetilde{Q} \| Q) \Big\} ,
\end{equation}
where
\begin{equation} \label{eqKL}
D(\widetilde{Q} \| Q) = \sum_x \widetilde{Q}(x) \log \bigg[ \frac{\widetilde{Q}(x)}{Q(x)} \bigg]
\end{equation}
denotes the divergence (Kullback-Leibler distance) between the
distributions $\widetilde{Q}$ and $Q$.
The relation (\ref{eqRelation2Gallager})
can be explained by averaging (\ref{eqExpMemoryless})
over the transmitted words ${\bf x}_m$,
and noting that the frequency of codewords of type $\widetilde{Q}$
in a random code generated by a distribution $Q$ is proportional to
$\exp \big\{-n D(\widetilde{Q} \| Q) \big\}$;
see \cite{CoverBook}.

\begin{figure}[ht]
\begin{center}
\epsfig{file=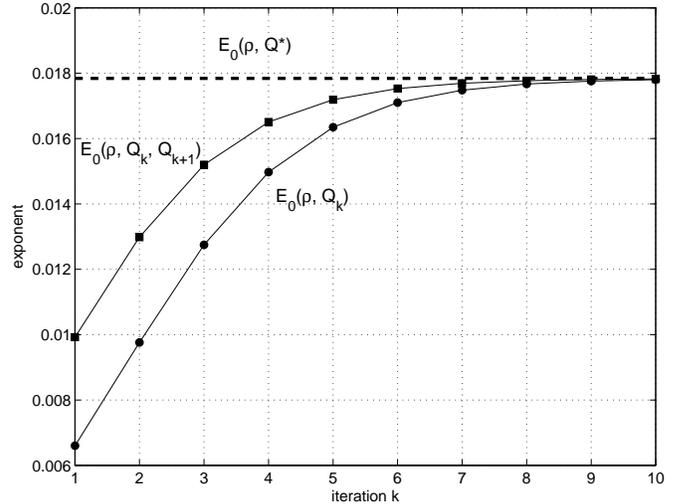, scale = 0.49}
\caption{Steps of the Arimoto algorithm.}
\label{figArimotoIteration}
\end{center}
\end{figure}

Figure~\ref{figArimotoIteration} shows the steps of the Arimoto algorithm
with respect to both the conditional and unconditional
Gallager functions, for a binary symmetric channel BSC(0.2), at $\rho=0.1$,
starting from an initial input distribution $Q_1=(0.1,0.9)$,
and converging to $Q^*=(0.5,0.5)$.

\subsection{{\bf Statement of Main Result}} \label{MainResult}
Let ${\{{{\bf X}\mathstrut}_{m}^{}\}\mathstrut}_{m\,=\,1}^{+\infty}$ be an infinite list of codewords of length $n$, selected independently with the probability measure $Q({\bf x})$.
This list can be viewed alternatively as a sequence of nested random codebooks of size $M$, ${{\cal C}\mathstrut}_{M}^{}\,\triangleq\,\{{{\bf X}\mathstrut}_{1}^{}, {{\bf X}\mathstrut}_{2}^{}, ...,  {{\bf X}\mathstrut}_{M}^{}\}$,
i.e. growing in size with each codeword.
Let ${Q\mathstrut}_{m}^{}$ denote the type of the codeword ${{\bf X}\mathstrut}_{m}^{}$.
We define the codeword with the maximum conditional Gallager exponent
in two steps:

1) In each codebook ${{\cal C}\mathstrut}_{M}^{}$ choose the codeword with the maximum
conditional Gallager exponent 
(\ref{eqExpMemoryless}), at $s=\frac{1}{1+\rho}\,$,
with respect to this codebook's size $M$.

2) Among all the codewords chosen in (1) (one for each $M \, = \, 1, 2, 3, ...$) choose the codeword which has the maximum such exponent,
which corresponds to
\begin{equation} \label{eqRandomIndex}
\max_{M}
\max_{m\,\leq\,M}
\bigg\{ E_{0}^{}\big(\rho, Q, {Q\mathstrut}_{m}^{}\big) \;
- \;\rho \, \frac{\log M}{n} \bigg\}.
\end{equation}

Note that implicit in the definition (\ref{eqRandomIndex}) is that each codeword competes at least with all its predecessors in the list,
which means that the conditional Gallager exponent is maximized only among the ``feasible'' codewords.
Denote the index of the codeword determined by (\ref{eqRandomIndex}) as ${N\mathstrut}_{n}^{}$.
Then the type ${Q\mathstrut}_{{N\mathstrut}_{n}^{}}^{}$ of the codeword ${{\bf X}\mathstrut}_{{N\mathstrut}_{n}^{}}^{}$ is close to ${Q\mathstrut}'$:


\vspace{3mm}

\begin{thm}[Favorite type]\label{prp} \hfill
\begin{equation} \label{eqProposition}
{Q\mathstrut}_{{N\mathstrut}_{n}^{}}^{}
\;\; \underset{n\,\rightarrow\,\infty}{\longrightarrow} \;\;
{Q\mathstrut}' \;\;\;\;\;\; {\rm in \;\; prob.}
\end{equation}
{\em where ${Q\mathstrut}'$ is the maximizing distribution (\ref{eqNewQ2}) in an iteration of the Arimoto algorithm.}
\end{thm}

\vspace{3mm}




\section{Proof of Main Result} \label{Proof}
The following lemma gives an alternative (optimization) meaning to the maximizing distribution (\ref{eqNewQ2}) in one iteration of the Arimoto algorithm.
As we shall see later, this lemma serves as a basis for our result.


\vspace{3mm}

\begin{lemma} [Standard optimization] \label{lemma3}
\begin{align}
{Q\mathstrut}'
& = \arg\max_{\widetilde{Q}}^{} \big\{E_{0}^{}(\rho, Q, \widetilde{Q}) - \rho \; D(\widetilde{Q} \| Q) \big\}
\; = \; \nonumber \\
& \arg\;\max_{\widetilde{Q}:\,D(\widetilde{Q} \| Q) \,\leq\, D({Q\mathstrut}' \| Q)}^{}
E_{0}^{}(\rho, Q, \widetilde{Q}) ,
\label{eqMaximizer}
\end{align}
{\em where ${Q\mathstrut}'$ is the maximizing distribution (\ref{eqNewQ2}) in an iteration of the Arimoto algorithm.}
\end{lemma}

\vspace{3mm}


Since $\rho$ and $Q$ are fixed,
in the remainder of this section we shall use the simplified notation
$E(P) \, \triangleq \, E_{0}^{}(\rho, Q, P)$,
and also ${R\mathstrut}_{M}^{} \, \triangleq \, \frac{1}{n}\log M$.
(The $P$ here stands for an arbitrary channel input, and should not be confused with
the channel transition distribution.)


\vspace{3mm}

\begin{proof}
Since $E(P)$ is a linear function of $P$ and $D(P \| Q)$ is a convex ($\cup$) function of $P$,
the following expression is a concave ($\cap$) function of $P$:
\begin{align}
&L(P) \; = \; \nonumber\\
&E(P) \; - \; \rho
\big(
D(P \| Q) -
D({Q\mathstrut}' \| Q)\big)
\; + \; \lambda\bigg[\sum_{x}^{}P(x) - 1\bigg],
\label{eqLagrangian3}
\end{align}
for any $\rho > 0$ and $\lambda$.
Differentiating (\ref{eqLagrangian3}) and equating the derivative to zero, we obtain the maximizing distribution of
the LHS of (\ref{eqMaximizer}):
\begin{multline}
\frac{\partial L}{\partial P(x)} \, = \,
-\log \sum_{y}^{}{P\mathstrut}_{}^{\frac{1}{1+\rho}}(y \,|\, x)\bigg[\sum_{x'}^{}Q(x'){P\mathstrut}_{}^{\frac{1}{1+\rho}}(y \,|\, x')\bigg]^{\rho}\\
\;-\;\rho \log \frac{P(x)}{Q(x)} \; - \; \rho \;+\; \lambda
\; = \; 0 \nonumber
\end{multline}
\begin{multline} 
\Rightarrow \;\;\;\;\;\;
{P\mathstrut}_{}^{*}(x) \; = \; \\
\frac{1}{K}\,Q(x)\bigg[\sum_{y}^{}{P\mathstrut}_{}^{\frac{1}{1+\rho}}(y \,|\, x)\bigg(\sum_{x'}^{}Q(x'){P\mathstrut}_{}^{\frac{1}{1+\rho}}(y \,|\, x')\bigg)^{\rho}\,\bigg]_{}^{-1/\rho} \\
\; = \; {Q\mathstrut}'(x).
\nonumber
\end{multline}
Observe that
${P\mathstrut}_{}^{*} \, = \, {Q\mathstrut}'$ is also the maximizer of $E(P)$
under the constraint $D(P \| Q) \leq
D({Q\mathstrut}' \| Q)$.
Indeed, suppose there exist a better distribution $\widetilde{P}$, such that $E(\widetilde{P})\,>\, E({P\mathstrut}_{}^{*})$
and $D(\widetilde{P} \| Q) \leq
D({Q\mathstrut}' \| Q)$, but then also $L(\widetilde{P}) \, > \, L({P\mathstrut}_{}^{*})$, since $\rho > 0$, which is a contradiction.
\end{proof}


The key idea of the proof of Theorem~\ref{prp}
is to compare upper and lower bounds on the maximum
\begin{equation} \label{eqMaximum}
\max_{m}^{}
\big\{ E\big({Q\mathstrut}_{m}^{}\big)
-\rho {R\mathstrut}_{m}^{} \big\},
\end{equation}
%
which is equivalent to the double maximum in (\ref{eqRandomIndex}), as shown below.
Note that (\ref{eqMaximum})
is a random variable, because the type of the $m$-th codeword is random.
As a result of the remaining lemmas, the upper bound on (\ref{eqMaximum}) becomes a function of the random type ${Q\mathstrut}_{{N\mathstrut}_{n}^{}}^{}$,
whereas the lower bound becomes a tight deterministic bound.
We start with a pair of bounds which are valid with probability $1$:


\vspace{3mm}

\begin{lemma} \label{lemma1}
\begin{equation} \label{eqSymmetricBounds}
E\big({Q\mathstrut}_{{N\mathstrut}_{n}^{}}^{}\big) -\rho
{R\mathstrut}_{{N\mathstrut}_{n}^{}}^{}
\;\; \geq \;\; \max_{\;m\,\leq \, {M\mathstrut}_{}^{}}^{}\big\{E\big({Q\mathstrut}_{m}^{}\big) \big\} -\rho {R\mathstrut}_{{M\mathstrut}_{}^{}}^{}\,,
\end{equation}
{\em for all $M$.}
\end{lemma}

\vspace{3mm}


\begin{proof}
\begin{align}
E\big({Q\mathstrut}_{{N\mathstrut}_{n}^{}}^{}\big) -\rho {R\mathstrut}_{{N\mathstrut}_{n}^{}}^{} \; & = \;
\max_{m}^{}\big\{E\big(Q_{m}^{}\big) -\rho {R\mathstrut}_{m}^{} \big\} \nonumber \\
\; & = \; \max_{m}^{}\;\max_{M\,\geq \, m}^{}\big\{E\big(Q_{m}^{}\big) -\rho {R\mathstrut}_{M}^{} \big\} 
\nonumber \\
& = \;
\max_{M}^{}\;\max_{m\,\leq \, M}^{}\big\{E\big(Q_{m}^{}\big) -\rho {R\mathstrut}_{M}^{} \big\} \nonumber \\
& \geq \; \max_{m\,\leq \, {M\mathstrut}_{}^{}}^{}\big\{E\big(Q_{m}^{}\big) \big\} -\rho {R\mathstrut}_{{M\mathstrut}_{}^{}}^{}\,,
\nonumber
\end{align}
for all $M$.
\end{proof}

Aiming at Lemma~\ref{lemma3}, in a somewhat symmetric manner we would like to replace
$\max_{m\,\leq\,M} E\big({Q\mathstrut}_{m}^{}\big)$ on the RHS of (\ref{eqSymmetricBounds}) with $E({Q\mathstrut}')$,
and ${R\mathstrut}_{{N\mathstrut}_{n}^{}}^{}$ on the LHS of (\ref{eqSymmetricBounds}) with
$D\big({Q\mathstrut}_{{N\mathstrut}_{n}^{}}^{} \|\, Q\big)$.
Both of these substitutions can be accomplished only "in probability", using properties of types, as the next two lemmas show.


\vspace{3mm}

\begin{lemma} \label{lemma2}
{\em Choose ${M\mathstrut}_{n}^{} \, = \, \lceil\exp\{n D({Q\mathstrut}' \| Q)\}\rceil$.
Then for any $\epsilon\, > \, 0$ and $n$ sufficiently large, with high probability the RHS of (\ref{eqSymmetricBounds}) is further bounded from below as}
\begin{multline} 
\max_{\;\;m\,\leq \, {M\mathstrut}_{n}^{}}^{}\big\{E\big(Q_{m}^{}\big)\big\} -\rho {R\mathstrut}_{{M\mathstrut}_{n}^{}}^{} \;\; \geq \;\;\\
\max_{P:\,D(P \| Q) \,\leq\, D({Q\mathstrut}' \| Q)}^{}\big\{E(P)\big\} -\rho {R\mathstrut}_{{M\mathstrut}_{n}^{}}^{} - \epsilon.
\nonumber
\end{multline}
\end{lemma}

\vspace{3mm}


\begin{proof}
Given a list of ${M\mathstrut}_{n}^{}$ words, generated independently with the probability measure $Q({\bf x})$,
the types $Q_{{\bf x}_{}^{}}^{}$ satisfying $D(Q_{{\bf x}_{}^{}}^{} \| Q) < D({Q\mathstrut}' \| Q)$ will be found in the list with high probability.
On the other hand, the types satisfying $D(Q_{{\bf x}_{}^{}}^{} \| Q) > D({Q\mathstrut}' \| Q)$ with high probability will not be found in the list.
If we choose the word ${\bf x}$ which has the highest $E(Q_{{\bf x}_{}^{}}^{})$ in the list, as the RHS of (\ref{eqSymmetricBounds}) suggests, this procedure translates into the following optimization problem:
\begin{displaymath} 
\max \, \{\,E(P)\,\} \;\;\;\;\;\; {\rm subject \;\; to} \;\;\;\;\;\; D(P \| Q) \leq D({Q\mathstrut}' \| Q).
\end{displaymath}
Equivalently, for any $\epsilon\, > \, 0$ and $n$ sufficiently large, with high probability
\begin{displaymath}
\max_{\;\;m\,\leq \, {M\mathstrut}_{n}^{}}^{}\big\{E\big(Q_{m}^{}\big)\big\} \;\; \geq \;\;
\max_{P:\,D(P \| Q) \,\leq\, D({Q\mathstrut}' \| Q)}^{}\big\{E(P)\big\} - \epsilon.
\end{displaymath}
\end{proof}


\vspace{3mm}

\begin{lemma} \label{lemma5}
{\em For any $\epsilon\, > \, 0$ and $n$ sufficiently large, with high probability
the LHS of (\ref{eqSymmetricBounds}) is further bounded from above as}
\begin{displaymath} 
E\big({Q\mathstrut}_{{N\mathstrut}_{n}^{}}^{}\big) -\rho {R\mathstrut}_{{N\mathstrut}_{n}^{}}^{} \; \leq \;
E\big({Q\mathstrut}_{{N\mathstrut}_{n}^{}}^{}\big) -\rho D\big({Q\mathstrut}_{{N\mathstrut}_{n}^{}}^{} \|\, Q\big) + \epsilon.
\end{displaymath}
\end{lemma}

\vspace{3mm}


\begin{proof}
Observe that
the codeword ${{\bf X}\mathstrut}_{{N\mathstrut}_{n}^{}}^{}$ has
the minimal index among the codewords of the type ${Q\mathstrut}_{{N\mathstrut}_{n}^{}}^{}$ in the list.
For any given type $Q_{{\bf x}_{}^{}}^{}$ define a random variable
\begin{displaymath}
{R\mathstrut}_{Q_{{\bf x}_{}^{}}^{}}^{} \; \triangleq \;
\min_{\;Q_{m}^{} \, = \, Q_{{\bf x}_{}^{}}^{}}\big\{{R\mathstrut}_{m}^{}\big\},
\end{displaymath}
which corresponds to the index of the first instance of the type $Q_{{\bf x}_{}^{}}^{}$ in the list.
Then
\begin{align}
& {\rm Pr}\left\{{R\mathstrut}_{{N\mathstrut}_{n}^{}}^{} \; \leq \;
D\big({Q\mathstrut}_{{N\mathstrut}_{n}^{}}^{} \|\, Q\big) - \epsilon\right\}
\nonumber \\
& \leq \;
{\rm Pr}\bigg\{\bigcup_{\;Q_{{\bf x}_{}^{}}^{}}\bigg[{R\mathstrut}_{Q_{{\bf x}_{}^{}}^{}}^{} \; \leq \;
D\big(Q_{{\bf x}_{}^{}}^{} \| Q\big) - \epsilon\bigg]\bigg\}
\nonumber \\
& \leq \;
\sum_{\;Q_{{\bf x}_{}^{}}^{}}
{\rm Pr}\big\{{R\mathstrut}_{Q_{{\bf x}_{}^{}}^{}}^{} \; \leq \;
D\big(Q_{{\bf x}_{}^{}}^{} \| Q\big) - \epsilon\big\}
\nonumber \\
& \leq \;
\sum_{\;Q_{{\bf x}_{}^{}}^{}}
{e\mathstrut}_{}^{\displaystyle n\big(D\big(Q_{{\bf x}_{}^{}}^{} \| Q\big) - \epsilon\big)}
{e\mathstrut}_{}^{\displaystyle -n D\big(Q_{{\bf x}_{}^{}}^{} \| Q\big)}
\nonumber \\
& = \;
\sum_{\;Q_{{\bf x}_{}^{}}^{}}
{e\mathstrut}_{}^{\displaystyle -n\epsilon}
\; = \; {e\mathstrut}_{}^{\displaystyle -n(\epsilon + o(1))}.
\nonumber
\end{align}
Which implies that with high probability
\begin{displaymath} 
{R\mathstrut}_{{N\mathstrut}_{n}^{}}^{} \; \geq \;
D\big({Q\mathstrut}_{{N\mathstrut}_{n}^{}}^{} \| \, Q\big) - \epsilon,
\end{displaymath}
and the lemma follows.
\end{proof}

Combining lemmas~\ref{lemma1},~\ref{lemma2}, and~\ref{lemma5}, we obtain
\begin{multline}
E\big({Q\mathstrut}_{{N\mathstrut}_{n}^{}}^{}\big) -\rho D\big({Q\mathstrut}_{{N\mathstrut}_{n}^{}}^{} \| \, Q\big)
\; \geq \;\\
\max_{P:\,D(P \| Q) \,\leq\, D({Q\mathstrut}' \| Q)}^{}\big\{E(P)\big\} -\rho {R\mathstrut}_{{M\mathstrut}_{n}^{}}^{} - 2\epsilon.
\label{eqAfterLemmas}
\end{multline}
Since
${R\mathstrut}_{{M\mathstrut}_{n}^{}}^{} \, \rightarrow \, D({Q\mathstrut}' \| Q)$ as $n \, \rightarrow \, \infty$,
by Lemma~\ref{lemma3}
\begin{multline}
\max_{P:\,D(P \| Q) \,\leq\, D({Q\mathstrut}' \| Q)}^{}\big\{E(P)\big\}
-\rho {R\mathstrut}_{{M\mathstrut}_{n}^{}}^{}
\;\; \underset{n\,\rightarrow\,\infty}{\longrightarrow} \;\;\\
\max_{P}^{}\big\{E(P) -\rho D(P \| Q) \big\}.
\nonumber
\end{multline}
Therefore, we can rewrite (\ref{eqAfterLemmas}) as
\begin{multline} 
E\big({Q\mathstrut}_{{N\mathstrut}_{n}^{}}^{}\big) -\rho D\big({Q\mathstrut}_{{N\mathstrut}_{n}^{}}^{} \| \, Q\big)
\; \geq \;\\
\max_{P}^{}\big\{E(P) -\rho D(P \| Q) \big\} \; - \; 2\epsilon.
\nonumber
\end{multline}
By continuity, applying Lemma~\ref{lemma3} again, we have (\ref{eqProposition}). \hfill $\Box$


\section{Discussion} \label{Discussion}
%
It follows from (\ref{eqRelation2Gallager}), (\ref{eqMaximizer}),
the non-negativity of the divergence \cite{CoverBook},
and since $Q^*_\rho$ is the unique fixed point of the Arimoto algorithm,
that
\begin{equation} \label{eqDiscussion1}
E_0\big(\rho,Q, {Q\mathstrut}'\big)
\geq
E_0\big(\rho,Q, Q\big)
\geq
E_0\big(\rho,Q\big) ,
\end{equation}
with equality if and only if $Q$ is the optimum input $Q^*_\rho$ in (\ref{eqQ*}).
In particular, it can be shown directly with the help of Jensen's inequality that
%
\begin{equation} \label{eqDiscussion2}
E_0\big(\rho,Q, {Q\mathstrut}'\big) \;
- \;\rho D({Q\mathstrut}' \| Q) \;\; \geq \;\; E_{0}^{}(\rho,\,Q) .
\end{equation}
Therefore, the type ${Q\mathstrut}'$ gives
a conditional error exponent (\ref{eqExpMemoryless}) which is higher than
the unconditional error exponent $E_{0}^{}(\rho,\,Q) - \rho R\;$ by a gap of at least $\rho D({Q\mathstrut}' \| Q)$.
Our results suggest the existence of an interesting phenomenon
that the lowest individual error probabilities in random codebooks occur not at the lowest rates,
but as a result of a trade-off between the probabilities of good types and the number of competing codewords.
When this trade-off is combined with the conditional Gallager exponent,
we obtain a better type for random codebook generation,
which surprisingly coincides with the outcome of the Arimoto algorithm and thereby guarantees convergence to the optimal
random codebook distribution, for a given channel.

One difficulty in such an input adaptation scheme,
is that it relies on channel knowledge,
since the conditional Gallager exponent assumes maximum likelihood decoding.
Furthermore, the optimum selection of the parameter $\rho$ depends on the channel (and the coding rate).
Another issue is that the system
requires estimation of the error exponent of each codeword, which is possible only after many transmissions.
And another weakness of the current analysis is that it assumes
working at a fixed rate $R < C$.  
So the current result is still far from our final goal of a universal adaptive scheme
for channel input adaptation, that approaches the true channel capacity $C$.
Some modification of the analysis is required to support a mechanism which is completely independent of the channel, and only relies on tentative estimates of codeword goodness at the receiver, and a feedback to the transmitter.  This future work will be reported elsewhere.


\section*{Appendix}
\subsection*{Proof of Lemma~\ref{lemma0}}
For memoryless $Q({\bf x})$ and ${P\mathstrut}_{}^{}({\bf y}\,|\,{\bf x})$ the bound (\ref{eqEnsemble}) becomes
\begin{align}
& {M\mathstrut}_{}^{\rho}\sum_{\bf y}{P\mathstrut}_{}^{1-s\rho}({\bf y}\,|\,{\bf x}_{m}^{})\bigg[\sum_{{\bf x}}^{}Q({\bf x}){P\mathstrut}_{}^{s}({\bf y}\,|\,{\bf x})\bigg]_{}^{\rho}
\nonumber \\
& = \;
{M\mathstrut}_{}^{\rho}\sum_{\bf y}{P\mathstrut}_{}^{1-s\rho}({\bf y}\,|\,{\bf x}_{m}^{})\bigg[\sum_{{\bf x}}^{}\prod_{k=1}^{n}Q({x\mathstrut}_{k}^{}){P\mathstrut}_{}^{s}({y\mathstrut}_{k}^{}\,|\,{x\mathstrut}_{k}^{})\bigg]_{}^{\rho} 
\nonumber
\end{align}
\begin{align}
& = \;
{M\mathstrut}_{}^{\rho}\sum_{\bf y}{P\mathstrut}_{}^{1-s\rho}({\bf y}\,|\,{\bf x}_{m}^{})\prod_{k=1}^{n}\bigg[\sum_{ x}^{}Q(x){P\mathstrut}_{}^{s}({y\mathstrut}_{k}^{}\,|\,x)\bigg]_{}^{\rho} 
\nonumber \\
& = \;
{M\mathstrut}_{}^{\rho}\sum_{\bf y}\prod_{k=1}^{n}{P\mathstrut}_{}^{1-s\rho}({y\mathstrut}_{k}^{}\,|\,{x\mathstrut}_{m, \,k}^{})\bigg[\sum_{ x}^{}Q(x){P\mathstrut}_{}^{s}({y\mathstrut}_{k}^{}\,|\,x)\bigg]_{}^{\rho} 
\nonumber \\
& = \;
{M\mathstrut}_{}^{\rho}\prod_{k=1}^{n}\sum_{y}{P\mathstrut}_{}^{1-s\rho}(y\,|\,{x\mathstrut}_{m, \,k}^{})\bigg[\sum_{ x}^{}Q(x){P\mathstrut}_{}^{s}(y\,|\,x)\bigg]_{}^{\rho} 
\nonumber \\
& = \;
{M\mathstrut}_{}^{\rho}\exp \Bigg\{-n\,\Bigg(\!-\frac{1}{n}\sum_{k\,=\,1}^{n}\log \sum_{y}{P\mathstrut}_{}^{1-s\rho}(y\,|\,{x\mathstrut}_{m, \,k}^{})
\nonumber \\
& \;\;\;\;\;\;\;\;\;\;\;\;\;\;\;\;\;\;\;\;\;\;\;\;\;\;\;\;\;\;\;\;\;\;\;\; \times\bigg[\sum_{ x}^{}Q(x){P\mathstrut}_{}^{s}(y\,|\,x)\bigg]_{}^{\rho} \Bigg)\Bigg\} 
\nonumber \\
& = \;
{M\mathstrut}_{}^{\rho}\exp \Bigg\{-n\,\Bigg(\!-\sum_{x}^{}Q_{{\bf x}_{m}^{}}^{}(x)\log \sum_{y}^{}{P\mathstrut}_{}^{1-s\rho}(y \,|\, x)
\nonumber \\
& \;\;\;\;\;\;\;\;\;\;\;\;\;\;\;\;\;\;\;\;\;\;\;\;\;\;\;\;\;\;\;\;\;\;\;\; \times\bigg[\sum_{x'}^{}Q(x'){P\mathstrut}_{}^{s}(y \,|\, x')\bigg]^{\rho} \Bigg)\Bigg\} 
\nonumber \\
& = \;
\exp \big\{-n\big[E_0 \big(s, \rho, Q, {Q\mathstrut}_{{\bf x}_{m}^{}}^{}\big) -\rho {R\mathstrut}_{}^{} \big] \big\}.
\nonumber
\end{align}
\hfill $\Box$

\section*{Acknowledgement}
We thank Yuval Kochman for interesting discussions in the beginning of this research,
and Nir Weinberger for helpful comments.
\bibliographystyle{IEEEtran}
\bibliography{randomexponent}

\end{document}